\documentclass[rmp,aps,nofootinbib]{revtex4}
\usepackage{graphicx}
\usepackage{dcolumn}
\usepackage{bm}

\begin{document}

\noindent {\Huge \bf Evolution in complex systems}

\vspace{2cm}

\noindent Paul Anderson$^1$, Henrik Jeldtoft Jensen$^{1*}$, L.P.Oliveira$^1$ and
Paolo Sibani$^{1,2,\dag}$

\vspace{2cm}

\noindent $^1$ Department of Mathematics, Imperial College, 180 Queen's Gate, London SW7 2AZ, U.K.\\
$^2$ Department of Physics, University of Oxford, 1 Keble Road, Oxford OX1 3NP, UK.\\
$^\dag$ Permanent address: Physics Department, University of Southern Denmark, 
Campusvej 55, \\ \hspace*{.3cm}DK-5230 Odense M, Denmark.

\vspace{2cm}

\noindent $^*$Author to whom correspondence should be addressed: h.jensen@imperial.ac.uk 

\vspace{2cm}



\begin{abstract}
What features characterise complex system dynamics? Power laws and scale 
invariance of fluctuations are often taken as the hallmarks of complexity, 
drawing on analogies with equilibrium critical phenomena[1-3]. Here we argue 
that slow, directed dynamics, during which the system's properties change 
significantly, is fundamental.  The underlying dynamics is related to a slow, 
decelerating but spasmodic release of an intrinsic strain or tension. 
Time series of a number of appropriate observables can be analysed to 
confirm this effect.  The strain arises from local frustration. As the 
strain is released through "quakes", some system variable undergoes record 
statistics with accompanying log-Poisson statistics for the quake event 
times[4]. We demonstrate these phenomena via two very different systems: 
a model of magnetic relaxation in type II superconductors and the Tangled 
Nature model of evolutionary ecology, and show how quantitative indications 
of ageing can be found.
\end{abstract}
\maketitle

Many macroscopic systems evolve through periods of relative quiescence separated 
by brief outbursts of hectic activity. We describe the prototype complex dynamics 
using two specific systems from physics and biology: the magnetic behaviour of 
type II superconductors and biological macroevolution. Each system is metastable 
when observed on short time scales, while at long time scales, each evolves 
towards greater stability. The models were introduced and discussed in general 
terms in Ref. 5-8. Our aim in the present paper is to focus on the nature of the 
long time relaxation associated with the intermittent activity.  This intermittent 
dynamics is in itself important and has attracted much interest[1,2]. Even more crucial 
is the often neglected fact that the punctuated dynamics of complex systems may lead 
to substantial changes in global properties, induced by the system following a 
distinct directed evolutionary path[9]. Descriptions borrowed from equilibrium and/or 
stationary systems are thus of limited value and will be unable to catch the essential 
time dependence of the dynamics. The long time effect of complex dynamics is evident 
in biological macroevolution, for example, in the form of a slowly decreasing 
extinction rate[10]. Similar effects are one of the main characteristics of spin 
glasses[11], and has been suggested to be relevant to the long time behaviour of 
geological faults[12,13] (though this non-stationary aspect of fault dynamics is 
often excluded in simple models[14]) and in economics[15]. 

Why do the properties of a complex system change as a result of the intrinsic system 
dynamics and how can this change be described quantitatively? In general terms, 
this is because complex systems consist of many components coupled together through 
a network of interactions. Since it is unlikely that the first, random configuration 
fully or even partly optimises all interactions, a complex system will initially be 
in a state of high frustration and strain, i.e. in a state unable to locally fulfil 
all constraints imposed by the mutual interactions in a many component system. 

The ensuing dynamics will act to release this strain and thus relax or optimise the 
system resulting in a more stable configuration. A many component system needs to 
find combined dynamical moves which collectively improve the distribution of 
interactions. Most of the dynamics of the individual degrees of freedom will not 
add up in a coherent and constructive way but will give rise to fluctuations about 
some metastable configuration. However, the inbuilt strain of the initial configuration 
does exert a directed push on all the components and will once in a while lead 
to coherent rearrangements of parts of the system. These essential events will 
be like snow avalanches or earthquakes in a geological fault. 
They induce an irreversible change in the properties of the system. We will 
call these events {\em quakes} to stress their dramatic effect on the stability 
of the system. 

It is important to distinguish the dynamics of complex systems presented here 
from the avalanche scenario outlined for Self-Organised Critical systems[1,2]. 
SOC emphasises that the concept of scale invariance or criticality (as encountered 
in equilibrium phase transitions) are of generic relevance to complex systems. 
The reason for this, as was argued in the seminal paper by Bak, Tang and 
Wiesenfeld (BTW)[16], is that large collections of interacting over-damped degrees 
of freedom which evolve according to dynamics controlled by thresholds will, if they 
are slowly driven by external actions, self-organise into a state which lacks any 
characteristic scale except for the one imposed by the finite system size. The 
widespread observation of power laws in nature is, within the SOC paradigm,
considered to be a consequence of the anticipated scale invariance of the 
stationary self-organised critical state. These power laws are the stationary 
probability distributions describing the SOC response in terms of (generalised)
avalanche events. Let us mention a few examples. In the BTW sandpile model[16], 
sand is sprinkled on to a surface at random. After a while, a stationary state 
is expected to develop in which the distribution of sand avalanches is described 
by a power law[17]. In the forest fire model[18], a stationary state is 
established by randomly growing trees at rate $p$ on the empty sites of a 
lattice and having lightning ignite a tree at a much smaller frequency $\nu$. 
The self-organised stationary state was expected to lead to a scale 
invariant distribution of the sizes of fires[19].

For completeness, we mention that Boettcher and Paczuski[20] have used the term 
ageing in a less restrictive way, than we do in this paper, in their study 
of the Bak-Sneppen model[21]. In this model, the return time of the activity 
of an individual avalanche depends on the age of that avalanche. This type
of ageing is unrelated to the phenomena we discuss here since it doesn't 
involve any change with time of physical properties.

\begin{figure}
\scalebox{0.5}[0.5]{\includegraphics{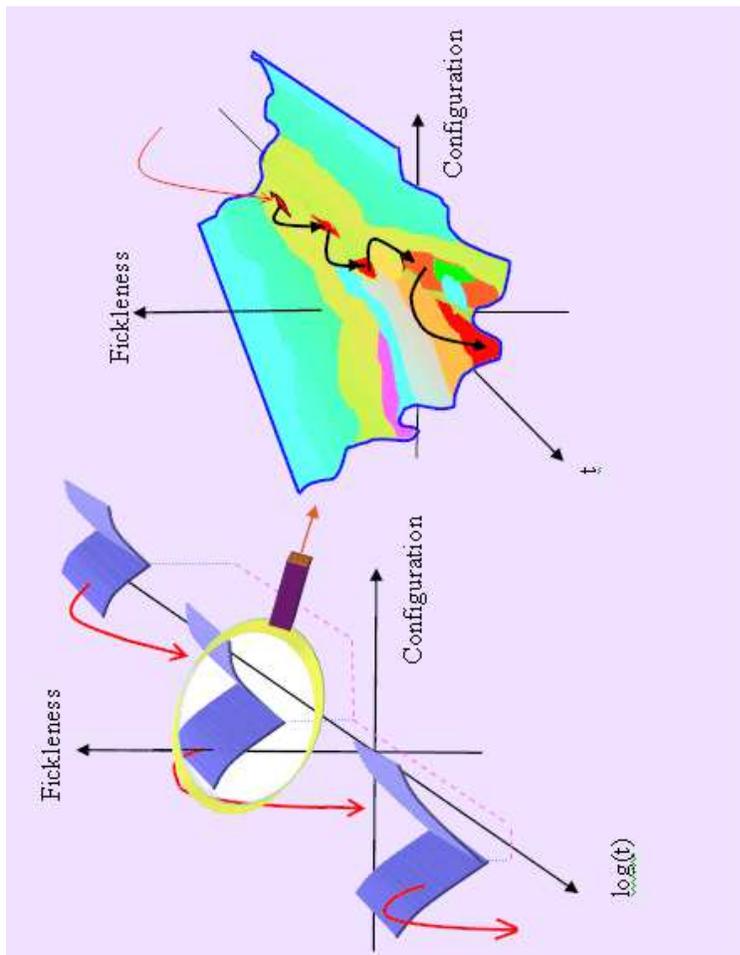}}
\caption{The dynamics of a complex system can be qualitatively summarised 
by considering the relation between time, configuration and fickleness.  
The smaller the fickleness value (i.e. the lower it is along the z-axis) 
the more stable the system becomes. The long time dynamics consists of a 
slow evolution in the form of jumps, or quakes, from one metastable configuration 
to the next, indicated by the sequence of ever deeper wells, or valleys, at 
the left of the figure. The quakes are only seen when the system is observed 
over many decades of time, hence the logarithmic time axis. The dynamics 
between the quakes is represented by the magnification shown on the right. 
On a linear (short) time scale, the system undergoes smaller jumps between 
sub-valleys within a single main valley. Short time dynamics slightly improves 
the stability of the system as indicated by the decrease of the system's 
fickleness with time. The quakes have a similar effect on a logarithmic time 
scale, as indicated by the deepening of the valleys on the left of the figure.
\label{fig1}}
\end{figure}

Whereas the focus in SOC is on the power laws encountered in the stationary 
state, the dynamics of complex systems we describe here concentrates on 
the fact that the quakes of complex systems gradually change both the 
physical and statistical properties of the system. We do not claim that 
complex systems are necessarily scale invariant and described by power 
laws, but we stress that the effect of a quake is to take the complex 
system into a new metastable configuration differing slightly from the 
previous state. Since the quake released some of the strain or improved 
a collection of interactions, this new metastable state will tend to be 
more stable. Hence, ever larger fluctuations are needed to take the system 
out of consecutive metastable configurations. Figure 1 illustrates the 
situation schematically. The system will tend to spend more and more time 
in the metastable states as it searches for a sufficiently large fluctuation 
that brings about a new and, on average, more stable configuration. This 
leads to a slowing down of the pace of evolution.  
To make this sketch more concrete and to be able to describe a methodology 
for observing and analysing the slow but crucial evolution of system 
properties, we now turn to a discussion of two specific systems. We 
choose two very different phenomena to illustrate the general nature 
of the description and analysis: magnetic relaxation in type II 
superconductors[5,6] and biological macroevolution[7,8]. A similar analysis 
can be carried out for spin glasses[22,23]. We emphasise that the analysis 
is non-intrusive and applicable to any system for which the time evolution 
of an appropriate observable is available.

The qualitative description of the previous paragraph suggests that the 
dynamics of complex systems tends to on average increase the stability, 
or decrease what we call in figure 1, fickleness. The transition from 
one metastable configuration to the next is accompanied by a drift 
in some measure: flux density for the model superconductor[5,6], population 
size for the model ecosystem[7,8] and energy in the spin glass model[22,23]. 
We call this measure the record parameter since its temporal evolution 
consists of a sequence of ever increasing record values. Each new record 
is triggered by a quake. On a logarithmic time scale, the quakes are 
essentially instantaneous and hence the quake number k is well determined 
by a single time, namely , the logarithm of the time of the onset of the 
quake. We now show how the dynamics is characterised by the statistical 
properties of the logarithmic waiting time between quakes, given by the sequence
$\tau_k=\ln(t_k)-\ln(t_{k-1})=\ln(t_k/t_{k-1})$.

We consider two models. Both are experimentally and observationally accessible[24-26]. 
First we consider the gradual penetration of the external magnetic field into 
the bulk of a type II superconductor after an initial ramping of the external 
magnetic field up to a fixed value. We use the Restricted Occupancy Model, or 
ROM[5,6], to simulate this system. The data presented here is for the more 
realistic three dimensional layered version of the ROM. We imagine a stack 
of two dimensional superconducting planes. Each plane is divided into $L\times
L$  squares. Here, $L=8$  and there are five layers, though similar behaviour 
is seen for different system sizes. A square can contain from zero up to a maximum number 
of $N_{c2}$ magnetic vortices. The state of the vortex system is specified 
by the number of vortices on each square. The vortices interact repulsively 
with the vortices in the nearest neighbour squares of the same plane 
(since parallel magnetic line segments repel each other). The only difference 
between the three dimensional and the two dimensional version of the ROM 
model is an attractive interaction between the vortices in squares right 
above each other in adjacent planes (an attraction similar to that between 
aligned compass needles). The model is updated using Monte Carlo dynamics[5,6]. 
The density of vortices at the boundary of the system represents the external 
magnetic field and is kept at a constant level. We monitor how the number 
of vortices inside the system increases with time as they enter the bulk 
sites of the model.

\begin{figure}
\scalebox{0.5}[0.5]{\includegraphics{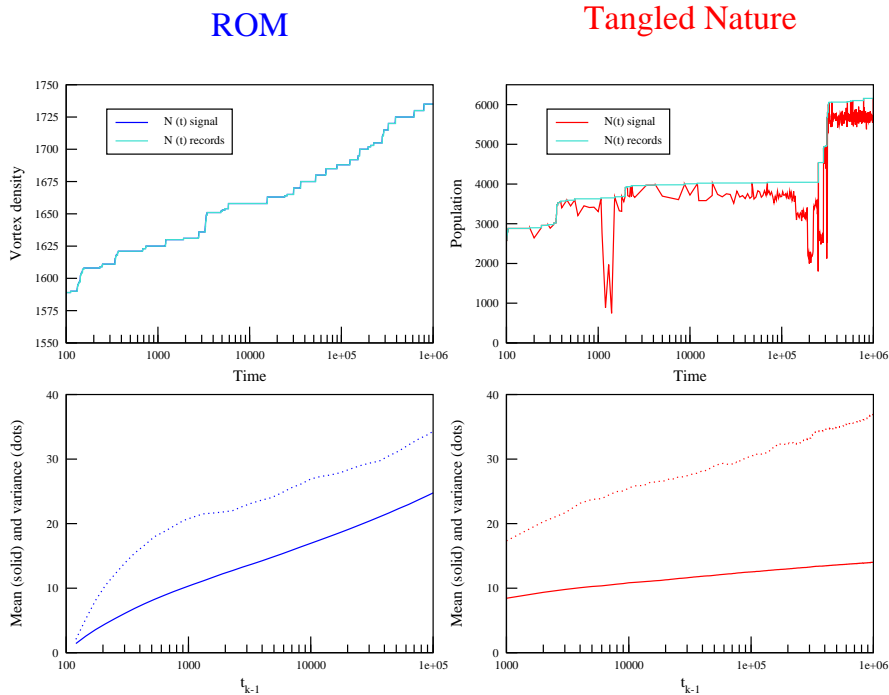}}
\caption{Temporal evolution in the ROM and the Tangled Nature model. 
The top plots show the value of $N(t)$ (and the corresponding records) 
for a single realisation. The lower panels show the average behaviour 
obtained from a set of realisations. The solid lines indicate the average 
number of quakes up to time t. The variance in the number of quakes is 
shown by the dotted curves. Both quantities exhibit an approximately 
linear dependence on the logarithm of time, as expected for a log-Poisson 
distribution. 
\label{fig2}}
\end{figure}

We obtain the sequence of quake waiting times, $\tau_k$  , from a time signal $N(t)$, 
which denotes the number of flux quanta inside the superconducting sample. 
As seen in figure 2, $N(t)$ is primarily a monotonically increasing step function. 
The characteristic behaviour of $N(t)$ can be identified from the observation that 
$N(t)$ is essentially equal to the record signal derived from it. The record is 
simply the largest value of $N(t)$ obtained up to time $t$. The jumps in $N(t)$  
define a sequence of quake times $t_k$  at which new flux quanta are able to 
penetrate into the sample. The intervals between the quake times are spent 
rearranging the internal flux in search of a configuration that better 
accommodates the magnetic pressure of the external applied field. The 
essential feature of the logarithmic slowing down of the evolution of 
the complex system's dynamics is modelled by a Poisson (or approximately 
Poisson) distribution of the logarithm of the quake times[4,22,23], i.e. a 
log-Poisson. This implies that the dynamics of the quakes are most naturally 
observed on a logarithmic time scale. Figure 2 confirms this, since the 
average and variance of the number of quakes of a set of independent 
realisations of $N(t)$ increases linearly with the logarithm of time, as 
expected for a log-Poisson process. The logarithmic time dependence is 
equivalent to a rate of events, $\nu$ , which decays inversely with 
time: $\nu\propto 1/t$. We note that the internal state at time $t$ is 
characterised by the highest number of flux quanta achieved up to that time. 

As our second example, we use the Tangled Nature model of evolutionary ecology[7,8]
to simulate the macroevolution of an ecosystem. This is an individual based model 
consisting of different interacting genotypes each characterised by a sequence of 
$L$ numbers which can be +1 or -1. This is to be thought of as the individual's 
genotype. For the results shown here, $L=20$, allowing for up to $2^{20}$  different 
genotypes, but very similar findings have been obtained for $L=8$. Reproduction 
is asexual and the reproduction probability of an individual is determined 
according to a weight function calculated from the frequency dependent interactions 
it has with other genotypes[7,8]. During reproduction, mutations can occur with a 
fixed probability. This leads to motion of the population in genotype space. Death 
consists of individuals being removed from the system with a fixed probability 
independent of time and genotype.

The time signal $N(t)$ is defined as the size of the entire population of the ecosystem.  
The choice of measure is not strictly unique. It is obviously important for the 
experimental or observational verification of the scenario outlined here that more 
than one choice exists for the signal $N(t)$ in general. For example, the records of 
the total number of occupied genotypes follow a very similar pattern to the total 
population. We see in figure 2 that the population size exhibits an overall increase 
with time, though fluctuations can make $N(t)$ decrease for short intervals. For 
a single realisation of $N(t)$, we derive the record signal, i.e. the largest value 
of $N(t)$ obtained up to time $t$. The quake times are identified as the record 
times of $N(t)$. The exact relation between these record times and the transitions 
between different metastable states of the Tangled Nature model is not clear 
but our numerical investigations gives us good reason to believe that the statistics 
of the record times mirrors the statistics of the transition times between metastable 
states. Averaged over a collection of independent realisations, $N(t)$ is found to 
increase logarithmically. Thus the record is a good indicator of the long time 
behaviour of $N(t)$. We expect the quakes occurring at these times (corresponding 
to reorganisation through species extinction and creation) to be responsible for 
the gradual collective adaptation observed in the model[7,8]. The quake times in 
this model are more difficult to identify compared to the ROM due to stronger fluctuations. 
Figure 2 shows that the records follow the same Poisson statistics on a logarithmic time 
scale as discussed above for the magnetic flux model. Again we see that the average 
and variance of the cumulated number of quakes increase linearly with the logarithm 
of time. The difference in the slopes might be attributed to a certain degree of 
over-counting. This may happen when a single quake is composed of a rapid succession 
of micro-quakes which may be counted as separate quakes, even though strictly they
are part of the same quake. The precise identification of quakes in Tangled Nature 
is a very difficult problem.

The constant logarithmic rate of quake events implies that the average waiting
time between quakes grows linearly with the age of the system. This implies 
that the variable $X=(t_k-t_{k-1})/t{k-1}$ should fluctuate about a constant value. 
This behaviour is to be distinguished from an ordinary Poisson process for which 
the average time between events is independent of time. In this case the variable 
$X$ will exhibit a rather rapid decay proportional to $1/T_{k-1}$. However, note that for a 
finite observation time $t_{obs}$, the ratio $X$ must decrease as $t_{k-1}$ approaches 
$t_{obs}$ since $t_k-t_{k-1}$ cannot exceed the value $t_{obs}- t_{k-1}$.

\begin{figure}
\scalebox{0.5}[0.5]{\includegraphics{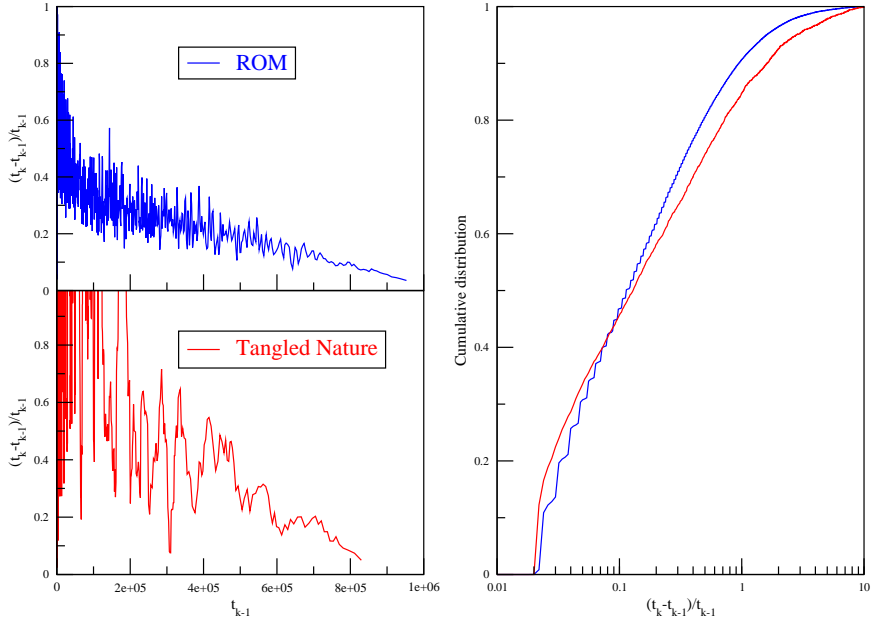}}
\caption{Evidence of ageing. The plots on the left show the ratio of the waiting 
time between quakes, $t_k-t_{k-1}$,  and the time of the $(k-1)$th quake $t_{k-1}$, 
as a function of $t_{k-1}$. For a stationary process, where the average value of the 
durations $t_k-t_{k-1}$ is a constant, $\Delta T$  say, this ratio will decay to 
zero like $\Delta t/t_{k-1}$ with increasing $t_{k-1}$. For the two models considered 
here, the ratio remains nearly constant over about six orders of magnitude. The 
slight decay at late times is due to the finite time $T_{obs}$ of the entire time 
sequence which imposes the constraint $t_k-t_{k-1}<T_{obs}-t_{k-1}$.  The panel 
to the right shows the cumulative distribution for this ratio in the two models. 
For a stationary process, this function would be a step function equal to one for 
$(t_k-t_{k-1})/t_{k-1}>0$.
} 
\label{fig}
\end{figure}

In figure 3, we show that for both models, $X$ varies no more than one order of 
magnitude even though $t_{k-1}$ spans six orders of magnitude. Thus, the older the 
system (equivalent to large values of $t_{k-1}$) the longer the time between records, 
$t_k-t_{k-1}$. The correlation functions shown in figure 4 indicate that 
correlations between consecutive quakes are negligible. We see that after 
a fast decay of the correlation function, a degree of negative (anti) 
correlation occurs before the correlations approach zero. It should also 
be noticed that these anti-correlations become more pronounced as the 
observation time $t_{obs}$  increases. (The largest  $t_{obs}$ in either case 
is one million: the number of time steps the simulations ran for.) The 
anti-correlation is due to the fact that, for a given observation 
window, a longer than average quiescent period will most likely be followed by a shorter one. 
This is because the activity slows down like $1/t_{k-1}$. It is accordingly 
impossible to achieve an observation window which is long compared to the 
longest waiting time and the numerically estimated correlation function is 
therefore always influenced by the finite duration of the observation no matter 
how big this window is taken to be. The effect can be easily mimicked in 
a standard Poisson process, but is not usually observed since under 
usual circumstances, the observation window can always be chosen to be 
much longer than the average time between the events.

\begin{figure}
\scalebox{0.5}[0.5]{\includegraphics{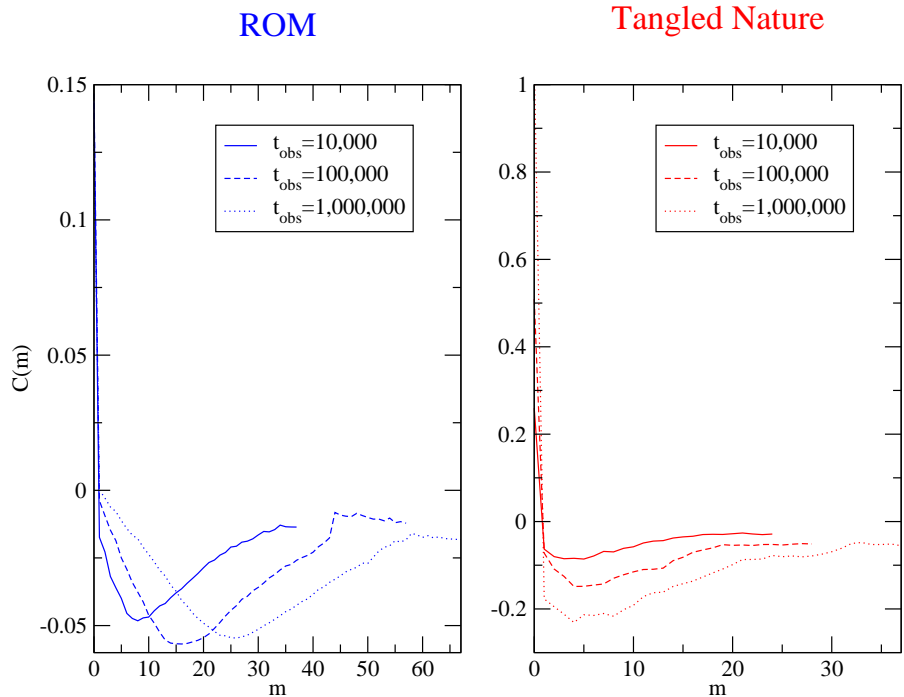}}
\caption{The plots show, as a function of $m$, the correlation between  
$\tau_k$ and  $\tau_{k+m}$, where $\tau_k=\ln(t_k)-\ln(t_{k-1}$ is the 
difference between the logarithmic times of occurrence of two successive quakes. 
The rapid decay of $C(m)$  agrees reasonably well with the theoretical form of 
a log-Poisson process, i.e. a Kronecker delta in $m$. The modest amount of 
anti-correlation seen for intermediate m values is due to observational 
effects, see main text. $t_{obs}$ is the observation time. For example, 
$t_{obs} =10,000$ means that only records that occur before $t=10,000$ are included 
in the analysis. Short range anti-correlations are observed for all window sizes.
} 
\label{fig4}
\end{figure} 

\begin{figure}
\scalebox{0.5}[0.5]{\includegraphics{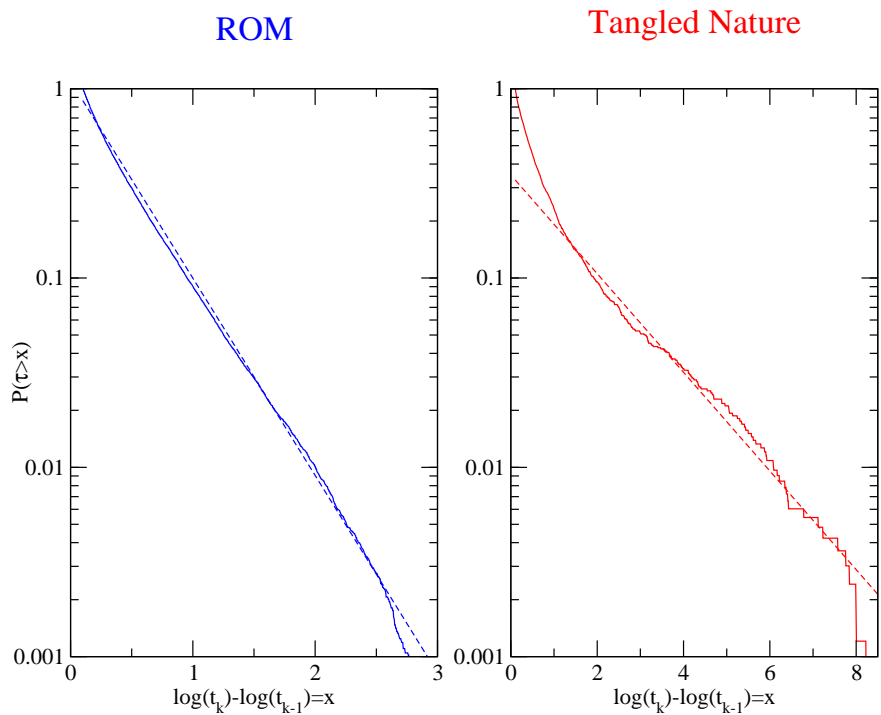}}
\caption{As further evidence that the records, or "quakes", follow the log-Poisson 
distribution, we include the distribution of the log waiting times for both the 
ROM and the Tangled Nature model, $P(\tau>x)$. For reference, we have shown an exponential 
distribution (broken line). If a series of records follows the log-Poisson distribution, 
the cumulated distribution of the log waiting times should follow an exponential[4]. 
The fit is good in both cases and provides a strong indication that the data 
does indeed obey a log-Poisson.
\label{fig5}}
\end{figure} 

The rapid decay of the correlation functions does not of course imply that 
the quakes are statistically independent, though it is consistent with 
assuming independence. If the quakes are independent, the logarithmic 
time intervals between quakes should be exponentially distributed. In 
figure 5 we show that this is the case, to a good approximation. We 
consider this finding as further indication that consecutive quake 
waiting times are essentially statistically independent.

It is worth mentioning that it is difficult to precisely identify the metastable 
states. The transition from metastable state to the next might be best described 
by the change in some measure characterising the stability of the configurations.
This might be an eigenvalue of the stability matrix associated with a set of effective 
evolution equations or, as in the simulation study of spin-glass relaxation in Ref. 23, 
from a very detailed analysis of the internal energy. In general one will have to rely 
on the measures which are accessible, though perhaps less than optimal. We have 
demonstrated here that record dynamics in complex systems can be analysed approximately 
even if all one can obtain is some simple macroscopic measurable quantity.  

Metastable systems of great complexity with huge numbers of interacting heterogeneous 
components are common throughout nature. It is crucial to realise that they are 
forever evolving at a decelerating pace towards configurations of greater stability 
and so concepts from equilibrium and stationary systems will only be of relevance 
over relatively short time scales. We have demonstrated above how ageing and 
record dynamics can be detected and described within an analytical framework.

\newpage
\noindent {\bf Acknowledgements:}\\ We are indebted to Andy Thomas, Dan Moore and Gunnar Pruessner 
for their support with the computations. Support from EPSRC, the Portuguese FCT, a 
visiting fellowship from EPSRC and financial support from the Danish SNF are gratefully 
acknowledged. We thank D. Sornette for directing us to some relevant literature.

\vspace{2cm}

\noindent {\bf\large References:}\\

\noindent 1. Bak, P., {\it How Nature Works. The science of self-organized criticality} (Oxford University Press, 1997).\\ 
2. Jensen, H.J., {\it Self-Organized Criticality. Emergent Complex Behavior in Physical and Biological Systems} 
(Cambridge University Press, 1998).\\
3. Kauffman, S., {\it At Home In The Universe. The Search for Laws of Self-organization and Complexity} 
(Oxford University Press, 1995).\\
4. Sibani, P. and Littlewood, P., Slow dynamics from noise adaptation.{\it Phys. Rev. Lett.}  {\bf
71}, 1482-1485 (1993).\\
5. Nicodemi, M. and H.J. Jensen, H.J.,  Equilibrium and off-equilibrium dynamics in a model for vortices 
in superconductors.{\it  Phys. Rev.} B {\bf 65}, art. no. 144517 (2002).
6. Jensen H.J. and Nicodemi M., Memory effects in response functions of driven vertex matter, 
{\it Europhys. Lett.} {\bf 57}, 348-354 (2002).\\
7. Christensen, K., Collobiano, S.A. di, Hall, M. and Jensen, H.J., Tangled Nature: A Model of Evolutionary 
Ecology. {\it J. of Theor. Biol}. {\bf 216}, 73-84 (2002).\\ 
8. Hall, M., Christensen, K., Collobiano, S.A. and Jensen, H.J., Time dependent extinction rate and 
species abundance in a tangled-nature model of biological evolution.{\it Phys. Rev. E} {\bf
66}, art. no. 011904 (2002).  \\
9. Ziemelis K. (ed.) Complex Systems - Nature Insight review, {\it Nature}{\bf 410}, 241-284 (2001).\\
10. Newman M.E.J., Sibani. P. Extinction, diversity and survivorship of taxa in the fossil record.
{\it Proc R Soc Lond B} {\bf 266}, 1593-1599 (1999).\\
11. Fischer K.H. and Hertz J. {\it Spin Glasses} (Cambridge University Press, 1991).\\
12. Lee M.W., Sornette D. and Knopoff L., Persistence and Quiescence of Seismicity on Fault 
Systems, {\it Phys. Rev. Lett.} {\bf 83}, 4219-4222 (1999). \\
13. Helmstetter A. and Sornette D., Diffusion of epicenters of earthquake aftershocks, Omori's law, 
and generalized continuous-time random walk models. {\it Phys. Rev. E.} {\bf
66} art. no. 061104 (2002).\\
14. Olami, Z., Feder, H.J.S. and Christensen, K., Self-organized criticality in a continuous, 
non-conservative cellular automaton modelling earthquakes. {\it Phys. Rev. Lett.}
{\bf 68}, 1244-1247 (1992).\\
15. Sornette D., Johansen A. and Bouchaud J.-P., Stock market crashes, Precursors and Replicas, 
{\it J.Phys.I France} {\bf 6}, 167-175 (1996).\\
16. Bak P., Tang C., Wiesenfeld K., Self-organized criticality: An explanation of the 1/f noise.  
{\it Phys. Rev. Lett}.{\bf 59}, 381 (1987).\\
17. Drossel B., Scaling behaviour of the Abelian sandpile model.  {\it
Phys. Rev. E}.{\bf 61}, R2168 (2000) discusses the deviations from the simple power law form 
first anticipated for the avalanche distribution.\\
18. Drossel B. and Schwabl F., Self-organized critical forest-fire model, {\it
Phys. Rev. Lett.} {\bf 69}, 1629 (1992).\\
19. Pruessner G. and Jensen H.J., Broken scaling in the forest-fire model, 
{\it Phys. Rev. E}. {\bf 65}, 056707 (2002) and Grassberger P., Critical Behaviour of the 
Drossel-Schwabl Forest Fire Model, {\it New J Phys}. {\bf 4}, art. no. 17 (2002). 
Both papers conclude that the simple power law scaling reported earlier is an artefact 
of small system sizes. For large systems, simple scaling doesn't apply.\\
20. Boettcher S. and Paczuski M., Aging in a Model of Self-Organized Criticality, {\it
Phys. Rev. Lett}. {\bf 79}, 889 (1997).\\
21. Bak P. and Sneppen K., Punctuated equilibrium and criticality in a simple model of 
evolution, {\it Phys. Rev. Lett}.{\bf 71}, 4083 (1993).\\
22. Sibani, P. and Dall, J., Log-Poisson statistics and full aging in glassy systems, 
{\it Europhys. Lett.}{\bf 64}, 8 (2003). \\
23. Dall, J. and Sibani, P., Exploring valleys of aging: the spin glass case, 
{\it Eur. Phys. J. B} {\bf 36}, 233 (2003). \\
24. Cooper, V. S., and R. E. Lenski, The population genetics of ecological specialization in evolving 
E. coli populations. {\it Nature} {\bf 407},736 (2000). 
25. Lenski, R. E.,  Phenotypic and genomic evolution during a 20,000-generation experiment with the 
bacterium Escherichia coli. {\it Plant Breeding Reviews} {\bf 24}, 225 (2004).
26. Cohen L.F. and Jensen H.J., Open questions in the magnetic behaviour of High Temperature superconductors, 
{\it Rep. Prog. Phys.}{\bf 60}, 1581 (1997).

\end{document}